\documentclass{article}
\usepackage[T1]{fontenc}
\usepackage[latin9]{inputenc}
\usepackage{color}
\usepackage{units}
\usepackage{amssymb}
\begin{document}

\title{Quantised inertia from relativity and the uncertainty principle.}

\author{M.E. McCulloch%
\thanks{Plymouth University, Plymouth, PL4 8AA. mike.mcculloch@plymouth.ac.uk%
}}
\maketitle
\begin{abstract}
It is shown here that if we assume that what is conserved in nature
is not simply mass-energy, but rather mass-energy plus the energy
uncertainty of the uncertainty principle, and if we also assume that
position uncertainty is reduced by the formation of relativistic horizons,
then the resulting increase of energy uncertainty is close to that
needed for a new model for inertial mass (MiHsC, quantised inertia)
which has been shown to predict galaxy rotation without dark matter
and cosmic acceleration without dark energy. The same principle can
also be used to model the inverse square law of gravity, and predicts
the mass of the electron.
\end{abstract}

\section{Introduction}

\textcolor{black}{Although special relativity and quantum mechanics
have been partially merged in quantum field theories, some aspects,
and general relativity and quantum mechanics are still incompatible.
For example, relativity is based on a smooth spacetime and demands
locality, whereas quantum mechanics is modelled using discrete particles
and quantum experiments seem to demand non-locality {[}1-5{]}.}

\textcolor{black}{In some instances it has been possible to combine
general relativity and quantum mechanics, at least partially, for
example {[}6{]} proposed that the event horizons caused by the strong
gravity within black holes would seperate pairs of particles produced
by the quantum vacuum, leaving one to fall into the black hole and
one to escape, giving rise to a new kind of radiation called Hawking
radiation that originates from a combination of relativity (curved
space) and quantum mechanics on a large scale. There is now some evidence
that at least analogues of this process occur {[}7{]}.}

\textcolor{black}{{[}8{]}, {[}9{]} and {[}10{]} showed that when an
object accelerates, say, to the left, an information horizon, very
like an event horizon, forms to its right since information which
is limited to the speed of light by relativity cannot now get to the
object from behind that horizon. They showed that this horizon can
seperate paired virtual particles in a similar way to a black hole
event horizon, leading to the production of acceleration-dependent
Unruh radiation. This conclusion is now generally accepted, but see
{[}11{]} for remaining controversies. It is possible that Unruh radiation
has already been observed {[}12{]}.}

\textcolor{black}{An early inspiring attempt to implicate quantum
mechanics and the zero point field in inertial mass was made by {[}13{]}.
However, they required an arbitrary cutoff to make their scheme work.
Also, {[}14{]} questioned whether Unruh radiation might account for
inertial-MoND (Modified Newtonian Dynamics), but concluded that Unruh
radiation was unlikely to be the cause of inertia because it was isotropic. }

\textcolor{black}{A new model for inertia was proposed by {[}15, 16{]}.
It is called Modified inertia by a Hubble-scale Casimir effect, MiHsC
or quantised inertia. This model assumes that the inertia of an object
is due to the Unruh radiation it sees when it accelerates. The relativistic
Rindler horizon that appears in the opposite direction to its acceleration
damps the Unruh radiation on that side of the object producing an
anisotropic radiation pressure that looks like inertial mass {[}16{]}.
So inertia arises in this model from the interplay of relativity (horizons)
and quantum mechanics (Unruh waves). Also, when accelerations are
extremely low the Unruh waves become very long and are also damped,
this time equally in all directions, by the Hubble horizon (Hubble-scale
Casimir effect) {[}15{]}. This leads to a new loss of inertia as accelerations
become tiny. So MiHsC modifies the standard inertial mass ($m$) to
a modified one ($m_{i}$) as follows:}

\textcolor{black}{
\begin{equation}
m_{i}=m\left(1-\frac{2c^{2}}{|a|\Theta}\right)
\end{equation}
}

\textcolor{black}{where c is the speed of light, $\Theta$ is the
diameter of the observable universe and '|$a$|' is the magnitude
of the acceleration of the object relative to surrounding matter.
Eq. 1 predicts that for terrestrial accelerations (eg: $9.8m/s^{2}$)
the second term in the bracket is tiny and standard inertia is recovered,
but in low acceleration environments, for example at the edges of
galaxies (when a is tiny) the second term in the bracket becomes larger
and the inertial mass decreases in a new way so that quantised inertia
(MiHsC) can explain galaxy rotation without the need for dark matter
{[}17{]} and cosmic acceleration without the need for dark energy
{[}15,18{]}. There are also anomalies seen in Solar system probes
{[}19{]} that can be explained by this model {[}15,20{]}. Quantised
inertia does not significantly affect the predictions of general relativity
for high accelerations and only becomes significant for very low accelerations
or upon a change in acceleration.}

\textcolor{black}{Similarly, applying quantum mechanics on a large
scale {[}21{]} derived Newtonian gravity from the uncertainty principle.
The main aim of this paper is to extend {[}21{]} and show that both
gravity and quantised inertia can be derived by allowing large-scale
dynamics or horizons to determine the position uncertainty in the
Heisenberg uncertainty principle, and allowing the resulting energy
uncertainty to become real.}

\section{\textcolor{black}{Gravity from Uncertainty}}

\textcolor{black}{Imagine there are two Planck masses orbiting each
other. With Planck masses, we are still, just, in the quantum realm,
Heisenberg's uncertainty principle applies to their mutual position
uncertainty ($\Delta x$) given by the distance between them, and
momentum ($\Delta p$), and the total uncertainty is twice that for
a single particle}

\textcolor{black}{
\begin{equation}
\Delta p\Delta x\sim\hbar
\end{equation}
}

\textcolor{black}{Now $E=pc$ so}

\textcolor{black}{
\begin{equation}
\Delta\bar{E}\Delta\bar{x}\sim\hbar c
\end{equation}
}

\textcolor{black}{If a bigger mass $M$ has $N$ Planck masses in
it, and another big mass m has n of them, then we can add up all the
possible interactions (all the various uncertainties: $\hbar c$)
between the various Planck masses}

\textcolor{black}{
\begin{equation}
\Delta\bar{E}\Delta\bar{x}=\sum_{i=1}^{N}\sum_{j=1}^{n}(\hbar c)_{ij}
\end{equation}
}

\textcolor{black}{The double summation on the right hand side is equal
to the number of Planck masses in mass m ($m/m_{P}$) times the number
in M ($M/m_{P}$), where $m_{P}$ is the reduced Planck mass, so}

\textcolor{black}{
\begin{equation}
\Delta\bar{E}=\frac{\hbar cmM}{m_{P}^{2}\Delta\bar{x}}
\end{equation}
}

\textcolor{black}{Now let us imagine that the Planck masses within
$m$ and $M$ are being buffeted from all sides by particles from
the zero point field and moving at random. The net effect, forgetting
horizons for a moment, will be zero. Sometimes random motion will
increase the distance between the two objects, $\Delta x$, so their
uncertainty in energy, $\Delta E$, decreases, and sometimes it will
decrease $\Delta x$, so the uncertainty in energy, $\Delta E$, will
increase. This latter event means that energy will suddenly be available
that wasn't before, extracted from the decrease in position uncertainty,
and if the objects continue to move together then more energy will
be released in this way allowing the motion to continue. What if we
assume that the sum of the kinetic energy and the energy uncertainty
is conserved?}

\textcolor{black}{
\begin{equation}
\nicefrac{1}{2}m(\Delta v)^{2}+\frac{\hbar cmM}{m_{P}^{2}\Delta\bar{x}}=constant
\end{equation}
}

\textcolor{black}{Differentiating}

\textcolor{black}{
\begin{equation}
m\Delta v\frac{d(\Delta v)}{dt}=\frac{\hbar cmM}{m_{P}^{2}\Delta x^{2}}\frac{d(\Delta x)}{dt}
\end{equation}
}

\textcolor{black}{Since the right-most fraction can be written as
$\Delta v$ we get}

\textcolor{black}{
\begin{equation}
m(\Delta a)=\frac{\hbar cmM}{m_{P}^{2}\Delta x^{2}}
\end{equation}
}

\textcolor{black}{Now we assume that $m(\Delta a)=F$ (force) and
that the uncertainty of the average position ($\triangle x$) is the
orbital radius $r$}

\textcolor{black}{
\begin{equation}
F\sim\frac{\hbar c}{m_{P}^{2}}\frac{mM}{r^{2}}
\end{equation}
}

\textcolor{black}{This looks like Newton's gravity law, and if we
insert the value of the Planck mass, for which the value of G must
be assumed, we get }

\textcolor{black}{
\begin{equation}
F=\frac{GMm}{r^{2}}
\end{equation}
The force required to drive the motion only becomes available for
objects moving closer together since this reduces $\Delta x$ and
increases $\Delta E$ (the inevitability of attraction was not discussed
in {[}21{]}). In this model, gravity is a process by which quantum
mechanics applies at this large scale and converts position uncertainty
to energy uncertainty, which shows up as an acceleration-dependent
heat (Unruh radiation) and so it satisfies the second law of thermodynamics:
increasing entropy. It has therefore been shown that Newton's gravity
law can be produced if a summation is made for all interactions between
masses equal to the Planck mass, but this requires an assumption of
the value of $G$ {[}21{]}.}

\section{\textcolor{black}{Quantised Inertia from Uncertainty}}

\textcolor{black}{Again, using Heisenberg's momentum-position uncertainty
principle we get}

\textcolor{black}{
\begin{equation}
\Delta p\Delta x\sim\hbar
\end{equation}
}

\textcolor{black}{Since $E=pc$ we can write}

\textcolor{black}{
\begin{equation}
\Delta E\Delta x\sim\hbar c
\end{equation}
}

\textcolor{black}{The energy uncertainty is then $\Delta E\sim\hbar c/\Delta x$.
The new proposal here is that if the particle in question accelerates
and a relativistic Rindler horizon forms then this destroys knowledge
of all positions beyond the horizon and decreases the uncertainty
in position $\Delta x$. From Eq. 12 we would then expect the uncertainty
in energy to go up. Now, as above we assume that what is conserved
in nature is not mass-energy, but rather mass-energy plus the energy
uncertainty identified above, as follows}

\textcolor{black}{
\begin{equation}
m_{1}c^{2}+\frac{\hbar c}{\Delta x_{1}}=m_{2}c^{2}+\frac{\hbar c}{\Delta x_{2}}
\end{equation}
}

\textcolor{black}{where the $m_{1}$ and $m_{2}$ are the initial
and final inertial masses and $\Delta x_{1}$ and $\Delta x_{2}$
are the initial and final positional uncertainties. Note that the
energy uncertainty terms are usually many orders of magnitude smaller
than the mass-energy terms. Rewriting we get}

\textcolor{black}{
\begin{equation}
m_{2}-m_{1}=dm=\frac{\hbar}{c}\left(\frac{1}{\Delta x_{2}}-\frac{1}{\Delta x_{1}}\right)
\end{equation}
}

\textcolor{black}{Now we can start to consider relativistic horizons.
For an minimally-accelerated object (a zero acceleration cannot exist
in MiHsC) the maximum uncertainty in position has to be due to the
cosmic horizon, and equal to the radius of the cosmos, so $\Delta x_{1}=\Theta/2$
so that}

\textcolor{black}{
\begin{equation}
dm=\frac{\hbar}{c}\left(\frac{1}{\Delta x_{2}}-\frac{2}{\Theta}\right)
\end{equation}
}

\textcolor{black}{If an object then is subjected to an acceleration,
a, then a Rindler horizon forms at a distance $d=c^{2}/a$ away. So
the new uncertainty in position is smaller $\Delta x_{2}=c^{2}/a$
so that}

\textcolor{black}{
\begin{equation}
dm=\frac{\hbar}{c}\left(\frac{a}{c^{2}}-\frac{2}{\Theta}\right)
\end{equation}
}

\textcolor{black}{Rearranging we get}

\textcolor{black}{
\begin{equation}
dm=\frac{\hbar a}{c^{3}}\left(1-\frac{2c^{2}}{a\Theta}\right)
\end{equation}
}

\textcolor{black}{Now an acceleration '$a$' is associated with Unruh
radiation of wavelength $\lambda$ where, using Unruh's expression
for the Unruh temerature $T=\hbar a/2\pi ck$ and Wien's law $T=\beta hc/k\lambda$
where $\beta=0.2,$ it follows that that $a=4\pi^{2}c^{2}\beta/\lambda$.
Also $E=hc/\lambda$. Using these to replace the '$a$' in the factor,
we get}

\textcolor{black}{
\begin{equation}
dm=\frac{\hbar}{c^{3}}\times\frac{4\pi^{2}\beta cE}{2\pi\hbar}\times\left(1-\frac{2c^{2}}{a\Theta}\right)
\end{equation}
}

\textcolor{black}{So that}

\textcolor{black}{
\begin{equation}
dm=\frac{4\pi^{2}\beta}{2\pi}\times\frac{E}{c^{2}}\times\left(1-\frac{2c^{2}}{a\Theta}\right)
\end{equation}
}

\textcolor{black}{Using $E=mc^{2}$ we get}

\textcolor{black}{
\begin{equation}
dm=2\pi\beta m\left(1-\frac{2c^{2}}{a\Theta}\right)
\end{equation}
}

\textcolor{black}{This is the same as Eq. 1, except for the initial
factor of $2\pi\beta\sim1.26$ which could be due to the crudity of
this model, which has treated the Rindler horizon as being a sphere
around the object whereas it is a more complex shape. The important
point is that Eqs. 1 and 20, by allowing quantum mechanics and relativity
to interact in this way, can model the observed anomalous galactic
rotation without dark matter {[}17{]} and the observed cosmic acceleration
without dark energy {[}15,18{]}.}

\section{\textcolor{black}{Applications}}

\subsection{\textcolor{black}{Particle masses}}

\textcolor{black}{An electron can be regarded as a photon that has
become confined to a particular orbit and so Eq. 14 can be used to
predict the mass-energy of the electron as follows}

\textcolor{black}{
\begin{equation}
dm=\frac{\hbar}{c}\left(\frac{1}{\Delta x_{2}}-\frac{1}{\Delta x_{1}}\right)
\end{equation}
}

\textcolor{black}{Initially the photon is confined to the cosmic scale
so $\Delta x_{1}=\Theta/2$ and it is known that for it to form an
electron it must have the Compton wavelength $\lambda_{C}=2.426\times10^{-12}m$
so}

\textcolor{black}{
\begin{equation}
dm=\frac{\hbar}{c}\left(\frac{1}{\lambda_{C}}-\frac{2}{\Theta}\right)
\end{equation}
}

\textcolor{black}{Neglecting the second term, which since $\Theta\sim10^{26}m$
is about 38 orders of magnitude smaller than the first, we get}

\textcolor{black}{
\begin{equation}
dm=\frac{\hbar}{c\lambda_{c}}=9.1\times10^{-31}kg
\end{equation}
}

\textcolor{black}{This is very close to the mass of the electron measured
in experiments. Similarly we can consider the protons and neutrons
which are confined to the nucleus of radius $r_{n}=1.75\times10^{-15}m$
(for hydrogen) so that}

\textcolor{black}{
\begin{equation}
dm=\frac{\hbar}{c}\left(\frac{1}{r_{n}}-\frac{2}{\Theta}\right)=1.3\times10^{-27}kg
\end{equation}
This is close to the observed masses of the proton and neutron which
are $1.67\times10^{-27}kg$. Equation 24 also predicts a small correction
to the proton mass given by the second term in the bracket, which
is about 41 orders of magnitude smaller than the first term in the
bracket.}

\textcolor{black}{If we use the Planck length $1.616\times10^{-35}m$
instead this gives}

\textcolor{black}{
\begin{equation}
dm=\frac{\hbar}{c}\left(\frac{1}{l_{P}}-\frac{2}{\Theta}\right)=1.4\times10^{-7}kg
\end{equation}
}

\textcolor{black}{This is close to the Planck mass, which is $2.2176\times10^{-8}kg$.
The agreement is very close if we use a scale of $2\pi l_{P}$}

\textcolor{black}{
\begin{equation}
dm=\frac{\hbar}{c}\left(\frac{1}{2\pi l_{P}}-\frac{2}{\Theta}\right)=2.2\times10^{-8}kg
\end{equation}
}

\textcolor{black}{Thus the assumption that what is conserved in nature
is not mass-energy as previously assumed, but mass-energy plus the
energy uncertainty and assuming the position uncertainty is determined
by relativistic horizons, allows the calculation of some particle
masses in this way as well as Newtonian gravity and quantised inertia
(MiHsC).}

\section{\textcolor{black}{Discussion}}

\textcolor{black}{These derivations can be explained more intuitively
as follows. For gravity: As the radius of an orbit decreases and so
the uncertainty in position decreases, then the momentum ($dp=Fdx/c$)
or force ($F$) on the orbiting body must increase, producing an inverse
square law. In the above gravitational derivation, the correct value
for the gravitational constant G can only be obtained when it is assumed
that the gravitational interaction occurs between whole multiples
of the Planck mass, but this last part of the derivation involves
some circular reasoning since the Planck mass is defined using the
value for G (this was not discussed in the precursor gravity paper,
{[}21{]}). This paper also builds on {[}21{]} by showing how this
formalism specifically implies attraction rather than repulsion (previously
it could have been either).}

\textcolor{black}{For inertia: as an object accelerates, a relativistic
Rindler horizon forms in the opposite direction. This curtails the
object's observable space and reduces its uncertainty in position.
The uncertainty principle then implies that the uncertainty in momentum
(or energy) must increase, and the energy released agrees (within
the uncertainty of the calculation) with the specific energy required
for quantised inertia (MiHsC) which allows the prediction of galaxy
rotation without dark matter and cosmic acceleration without dark
energy.}

\section{\textcolor{black}{Conclusion}}

\textcolor{black}{The uncertainty principle of quantum mechanics states
that if the uncertainty in position reduces, then the uncertainty
in momentum increases. Relativity predicts that if an object accelerates,
a Rindler horizon forms, curtailing its observable space.}

\textcolor{black}{If we combine these two principles, the formation
of the Rindler horizon reduces position uncertainty, increasing energy
uncertainty. It has already been shown, in a similar way, that if
we accept this energy as being real, Newtonian gravity is the result,
though a value for G has to be assumed.}

\textcolor{black}{It is shown here that using the same method, the
model known as quantised inertia or MiHsC can also be derived, solving
the problems of galaxy rotation and cosmic acceleration, and predicting
the electron mass.}

\section*{\textcolor{black}{References}}

\textcolor{black}{{[}1{]} Einstein, A., B. Podolsky and N. Rosen,
1935. Can quantum mechanical desriptions of reality be considered
complete? Physical Review, 41, 777.}

\textcolor{black}{{[}2{]} Bell, J.S., 1964. On the Einstein-Podolsky-Rosen
paradox. 1, 3, 195.}

\textcolor{black}{{[}3{]} Aspect, A., Dalibard, G. Roger, 1982. Experimental
test of Bell's inequalities using time-varying analyzers. Phys. Rev.
Letters, 49, 25, 1804.}

\textcolor{black}{{[}4{]} Iorio L., 2015. Editorial for the Special
Issue 100 Years of Chronogeometrodynamics: The Status of the Einstein's
Theory of Gravitation in Its Centennial Year, Universe, vol. 1, no.
1, pp. 38-81, 2015}

\textcolor{black}{{[}5{]} Rovelli, C., Quantum gravity, Scholarpedia,
3(5), 7117.}

\textcolor{black}{{[}6{]} Hawking, S.W., 1975. Commun. Math. Phys.,
43, 199.}

\textcolor{black}{{[}7{]} Steinhauer, J., 2016. Observation of quantum
Hawking radiation and its entanglement in an analogue black hole.
}\textit{\textcolor{black}{Nature Physics}}\textcolor{black}{. http://dx.doi.org/10.1038/nphys3863}

\textcolor{black}{{[}8{]} Fulling, S.A., 1973. Nonuniqueness of canonical
field quantization in Riemannian space-time. Phys. Rev D., 7, 10,
2850.}

\textcolor{black}{{[}9{]} Davies, P.C.W., 1975. Scalar production
in Schwarzschild and Rindler metrics. J. Physics., 8 (4), 609.}

\textcolor{black}{{[}10{]} Unruh, W.G., 1976. Notes on black hole
explosions. Phys. Rev., D., 14, 870.}

\textcolor{black}{{[}11{]} Fulling, S.A., and G.E.A. Matsas, 2014.
Scholarpedia, 9(10): 31789.}

\textcolor{black}{{[}12{]} Smolyaninov, I.I., 1976. Photoluminescence
from a gold nanotip in an accelerated reference frame. Phys. Rev.
D., 14, 870.}

\textcolor{black}{{[}13{]} Haisch, B., Rueda A., and H.E. Puthoff,
1994. Phys. Rev. A., 49, 678.}

\textcolor{black}{{[}14{]} Milgrom, M., 1999. Phys. Lett. A., 253,
273.}

\textcolor{black}{{[}15{]} McCulloch, M.E., 2007. Modelling the Pioneer
anomaly as modified inertia. }\textit{\textcolor{black}{MNRAS}}\textcolor{black}{,
376, 338-342.}

\textcolor{black}{{[}16{]} McCulloch, M.E., 2013. Inertia from an
asymmetric Casimir effect, EPL, 101, 59001.}

\textcolor{black}{{[}17{]} McCulloch, M.E., 2012. Testing quantised
inertia on galactic scales. }\textit{\textcolor{black}{ApSS}}\textcolor{black}{,
Vol. 342, No. 2, 575-578.}

\textcolor{black}{{[}18{]} McCulloch, M.E., 2010. Minimum accelerations
from quantised inertia. }\textit{\textcolor{black}{EPL}}\textcolor{black}{,
90, 29001.}

\textcolor{black}{{[}19{]} Iorio, L., 2015. Gravitational anomalies
in the solar system? Int. J. Mod. phys. D, Vol. 24, id. 1530015}

\textcolor{black}{{[}20{]} McCulloch, M.E., 2008. Modelling the flyby
anomalies using a modification of inertia. MNRAS, 389, L57-60.}

\textcolor{black}{{[}21{]} McCulloch, M.E., 2014. Gravity from the
uncertainty principle. }\textit{\textcolor{black}{ApSS}}\textcolor{black}{,
Vol. 389, No. 2, 957-959.}
\end{document}